\documentclass[aps,superscriptaddress,showpacs]{revtex4}

\usepackage{amssymb}

\usepackage{amsmath}

\newcommand{\beq}{\begin{equation}}
\newcommand{\nec}{\newcommand}
\nec{\cts}{conformal transformations }
\nec{\fourg}{g_{\alpha\beta}} 
\nec{\grad}{\bigtriangledown}
\nec{\fourr}{^{(4)}R}
\nec{\detg}{^{(4)}g}
\nec{\eins}{\Biggl(R_{\alpha\beta} - \frac{1}{2}g_{\alpha\beta}R\Biggr)}
\nec{\kk}{K^{ab}K_{ab}}
\nec{\bec}{\begin{center}}
\nec{\eec}{\end{center}}
\nec{\bb}{B^{ab}B_{ab}}
\nec{\rp}{R - 8\frac{\grad^{2}\psi}{\psi}}
\nec{\pipi}{\pi^{ab}\pi_{ab}}
\nec{\beqq}{\begin{equation*}}
\nec{\eeqq}{\end{equation*}}
\nec{\V}{V(\psi)}
\nec{\delg}{\frac{\partial g_{ab}}{\partial t}}
\nec{\wt}{\widetilde}
\nec{\gt}{\longrightarrow}
\nec{\wh}{\widehat}

\begin{document}

\title{The physical gravitational degrees of freedom}

\author{E. Anderson}
\email{ea212@hermes.cam.ac.uk}
\affiliation{Astronomy Unit, School of Mathematical Sciences, 
Queen Mary University of London, 
London, E1 4NS, UK}
\affiliation{Department of Physics, P-412 Avadh Bhatia Physics 
Laboratory, University of Alberta, Edmonton Canada}
\affiliation {Peterhouse, Cambridge CB2 1RD, UK}
\affiliation{DAMTP, Centre for Mathemetical Sciences, Wilberforce Road 
Cambridge CB3 0WA, UK}

\author{J. Barbour}
\email{julian@platonia.com}
\affiliation{College Farm, South Newington, Banbury 0X15 4JG, UK}

\author{B.Z. Foster}
\email{bzf@physics.umd.edu}
\affiliation{Physics Department, University of Maryland, College Park,  
MD, USA}

\author{B. Kelleher}
\email{kelleher@physics.ubc.ca}
\affiliation{Physics Department, University College Cork, Ireland}
\affiliation{Department of Physics and Astronomy, University of British Columbia, Vancouver, BC V6T 1Z1, Canada}

\author{N. \'{O} Murchadha}
\email{niall@ucc.ie}
\affiliation{Physics Department, University College Cork, Ireland}

\date{\today}

\begin{abstract}
When constructing general relativity (GR), Einstein required 4D
general covariance. In contrast, we derive GR (in the compact, without boundary case) as a theory of evolving 3-dimensional conformal Riemannian geometries obtained by
imposing two general principles: 1) time is derived from change; 2)
motion and size are relative. We write down an explicit action based on
them. We obtain not only GR in the CMC gauge, in its Hamiltonian 3 + 1
reformulation but also all the equations used in York's conformal
technique for solving the initial-value problem. This shows that the
independent gravitational degrees of freedom obtained by York do not
arise from a gauge fixing but from hitherto unrecognized fundamental
symmetry principles. They can therefore be identified as the long-sought
Hamiltonian physical gravitational degrees of freedom. 
\end{abstract}


\maketitle

Since Einstein created GR 4D spacetime covariance has been taken 
as its axiomatic basis. However, much work has been
done in a dynamical approach that uses the 3+1 split into space and time
of Arnowitt, Deser, and Misner (ADM)\cite{adm}. This work has been
stimulated by the needs of astrophysics (especially gravitational-wave
research) and by the desire to find a canonical version of GR suitable
for quantization.

    The ADM formalism describes constrained Hamiltonian evolution of 3D 
spacelike hypersurfaces embedded in 4D spacetime. The intrinsic geometry
of the hypersurfaces is represented by a Riemannian 3-metric $g_{ij}$,
which is the ADM canonical coordinate. The corresponding canonical
momentum $\pi^{ij}$ is related to the extrinsic curvature $\kappa^{ij}$
of the embedding of the hypersurfaces in spacetime by $\pi^{ij}=-\sqrt
g(\kappa^{ij}-g^{ij}\kappa)$.

   The ADM dynamics, which respects full relativity of simultaneity by 
allowing free choice of the 3+1 split, is driven by two constraints. The
linear {\it{momentum constraint}} ${\pi^{ij}}_{;j}=0$ reflects the gauge
symmetry under 3D diffeomorphisms and is well understood. When it has
been quotiented out, the $3\times 3$ symmetric matrix $g_{ij}$ has three
degrees of freedom. The quadratic {\it{Hamiltonian constraint}}
$gH=-\pi^{ij}\pi_{ij}+\pi^2/2+gR=0$ reflects the relativity of
simultaneity -- the time coordinate can be freely chosen at each space
point. It shows that $g_{ij}$ has only two physical degrees of freedom.
The problem is to find them. The solution, if it exists, will break 4D
covariance.

    An important clue was obtained by York \cite{yor}, who perfected
Lichnerowicz's conformal technique \cite{Lich} for finding initial data
that satisfy the initial-value constraints of GR. In the Hamiltonian
formalism, these are the ADM Hamiltonian and momentum constraints.
Finding such data is far from trivial. York's is the only known effective
method. He divides the 6 degrees of freedom in the 3-metric into three
groups; 3 are mere coordinate freedoms, 1 is a scale part (a conformal
factor), and the two remaining parts represent the conformal geometry,
the `shape of space'. York's method also
introduces a distinguished foliation of spacetime -- and with it a
definition of simultaneity -- by hypersurfaces of constant mean
(extrinsic) curvature (CMC).

    Based on his success, York concluded that the two conformal degrees of 
freedom at each space point do represent the gravitational degrees of
freedom. If space is compact without boundary (CWB), the local shapes
interact with a solitary global variable -- the changing value of
the CMC. However, the view that the local shapes are the true degrees of
freedom has not been widely accepted. This is probably because York's
method has not hitherto been derived from a physical first principle but
on the basis of a mathematical condition: conformally invariant
decoupling of the momentum and Hamiltonian constraints. If one is to
break the gauge symmetry of 4D covariance to obtain true gravitational
degrees of freedom, a plausible alternative gauge symmetry must be found
and used to derive York's method.

    We show that it is relativity of scale (size). Adopting it in 
conjunction with the principles that time is derived from change and that
motion is relative, we recover the ADM Hamiltonian formalism but with
York's distinguished CMC foliation and all the equations of his method.
We give a physical explanation for the success of his method. His
physical degrees of freedom are unambiguously identified.

The configuration space on which we work is defined by successive 
quotienting. The space of suitably continuous Riemannian 3-metrics
$g_{ij}$ on a CWB 3-manifold is Riem. Its quotient wrt 3D diffeomorphisms
is {\it{superspace}}. The quotient of superspace wrt 3D conformal
transformations $g_{ij}\rightarrow \phi^4g_{ij}$ is {\it{conformal
superspace}} (CS) \cite{CS}, which York uses to parametrize initial data
that satisfy the ADM constraints. We define our theory on the marginally
larger space CS+V obtained by quotienting superspace wrt
{\it{volume-preserving}} 3D conformal transformations. The infinitely
many local scales are all relative but the overall scale is absolute.
This weak relaxation of full scale invariance means that the changing
total volume of space (the Lagrangian counterpart of York's
momentum-like CMC value) is one extra degree of freedom along with the
pure shape variables. It allows expansion of the universe.

To implement our Principle 1, i.e., that time is derived from change, we
construct on CS+V a theory of  `timeless geodesics',  parametrized by a
freely chosen label
$\lambda$. Local proper time is emergent in this  approach. Our Principle
2, i.e., that motion and size are relative, is implemented by invariance
under $\lambda$-dependent  3D diffeomorphisms and volume-preserving conformal transformations.

We achieve our aims by modifying the Baierlein--Sharp--Wheeler~(BSW)~\cite{bsw} form of the
GR action:
\beq 
    I_{\mbox{\scriptsize BSW\normalsize}} 
    = \int\textrm d\lambda
        \mathcal L_{\mbox{\scriptsize BSW\normalsize}}
    =\int {\textrm d}
    \lambda{\textrm d}^3x \sqrt{g}\sqrt{R}\sqrt{T},
\label{1}  
\end{equation} 
where $g=\textrm{det} (g_{ij})$, $R$ is the 3-d scalar curvature, and 
\beq
T=G^{ijkl}\frac{d g_{ij}}{d\lambda}\frac{d g_{kl}}{d\lambda} 
    \equiv G^{ijkl}(\dot{g}_{ij}-\pounds_{\xi} g_{ij})(\dot{g}_{kl} -
\pounds_{\xi}g_{kl}).
\end{equation}
Here 
$G^{ijkl} \equiv (g^{ik}g^{jl} - g^{ij}g^{kl})$ is the inverse DeWitt
supermetric,  and $\pounds_{\xi}g_{ij}= \nabla_{i}\xi_{j} +
\nabla_j\xi_i$ is the Lie derivative of the metric with respect to $\xi^i$. This action is invariant under local reparametrization of $\lambda$,  satisfying Principle 1.  The emergent proper time $t$ is related to the label  $\lambda$ by $\delta t=N\delta\lambda$, for $N \equiv \frac{1}{2}\sqrt{T/R}$. $\xi_i$ is effectively a gauge auxiliary that renders the action invariant  under 3-diffeomorphisms, satisfying the `motion is relative' part of Principle  2.  

The BSW action is defined on curves in superspace.  We considered 
generalizations of this action  in~\cite{rwr}.  As this led us almost uniquely back to the BSW action, it is a new derivation of GR.  In~\cite{bom}, we extended the BSW action to CS, satisfying our new (in gravitational theory) `size is relative' principle. While this led to a consistent theory, it is not GR.  Interestingly, in the asymptotically flat case, BSW on CS does generate GR, in the maximal guage. Here, we  extend  the BSW action only to CS+V. 
The volume-preserving restriction allows us to recover GR in the  York picture.  

We will denote a volume-preserving conformal transformation~(VPCT) with a
`hat': 
\beq
    g_{ij}(x) \rightarrow \wh{\omega}(x)^4g_{ij}(x).
\end{equation}  
One can construct a
VPCT  $\wh\omega$ from any unrestricted conformal transformation $\omega$:
\beq
    \wh{\omega} = \frac{\omega}{<\omega^6>^{1/6}},   \label{VPCT}
\end{equation}
where 
\beq
    <F> = \frac{\int\textrm{d}^3x\sqrt{g}\,F}
        {\int \textrm{d}^3x\sqrt{g}},
\end{equation} 
denotes the global average of some function $F$.  
One can express any VPCT $\wh{\omega}$ in this manner.

We implement the conformal symmetry in the BSW action~\eqref{1} by introducing an
auxiliary scalar  field $\phi$, constructing from it in the manner above
the field
$\wh\phi$.  Under a VPCT~(\ref{VPCT}),  we declare that $\wh\phi$
transforms as
\beq
   \wh\phi \rightarrow \frac{\wh\phi}{\wh\omega}.
\label{y}
\end{equation}
Consequently, the `corrected coordinates' $\overline g_{ij}\equiv\wh\phi^4 g_{ij}$
are invariant  under a VPCT.  We then re-express the BSW action in terms
of $\overline g_{ij}$,  obtaining an action functional in $g_{ij}$ and
$\phi$.  We vary $\phi$ freely, not $\wh\phi$.

On the introduction of the auxiliary variable $\phi$, the constituent parts of the BSW
action become
\beq 
    R\rightarrow \wh\phi^{-4}\,
    \biggl(R 
        - 8\frac{\grad^{2}\wh{\phi}}{\wh{\phi}}\biggr),
    \hskip 1cm T \rightarrow \wh{T} 
    = \wh{\phi}^{-8}G^{ijkl}
        \frac{d\wh{\phi}^{4}g_{ij}}{d\lambda}
       \frac{d\wh{\phi}^{4}g_{kl}}{d\lambda}, 
\end{equation}
with
\beq 
    \frac{d\wh{\phi}^{4}g_{ij}}{d\lambda} 
    =  \wh{\phi}^{4}\left[\dot{g}_{ij} 
        - \pounds_{\xi}g_{ij} 
        + \frac{4}{\wh{\phi}}(\dot{\wh{\phi}} 
        - \pounds_{\xi}\wh{\phi})g_{ij}\right].
\label{straight}
\end{equation} 
The action~\eqref{1} becomes 
\beq 
    I_{\textrm{CS+V}} 
    = \int\textrm{d}\lambda \int\textrm{d}^3x\sqrt{g}\,
    \wh\phi^{4}\,
    \sqrt{R - 8\frac{\grad^{2}\wh\phi}{\wh{\phi}}}
    \sqrt{\wh{T}}.
\label{action} 
\end{equation}

We now explain the rules by which we vary the action. A curve through the
configuration space  specifies a sequence $g_{ij}(\lambda)$ through Riem,
combined with a  $\lambda$-dependent conformal  factor $\wh\phi(\lambda)$ and a $\lambda$-dependent coordinate transformation $\chi_i(\lambda)$ (for  which $\xi_i = \partial \chi_i/\partial\lambda$).  We identify $\phi$ and $\chi_i$ as gauge variables,  so they need not reduce to the identity at the end points of the curve.  Therefore, we extremize the  action under fixed-end-point variations of $g_{ij}$ and free-end-point
variations of $\phi$ and   $\chi_i$.

We now illustrate free-end-point variation in a simple scale-invariant
$N$-particle model  \cite{jb}. The corrected coordinates are $\tilde q_{(i)}=aq_{(i)}$ and the corrected  velocities are  $\dot{\tilde q}_{(i)}=\dot aq_{(i)}+a\dot q_{(i)}$, 
${\mathcal L}={\mathcal L}(\tilde{q}_{(i)},\dot{\tilde q}_{(i)})$. 
Here $a$ is the scaling auxiliary and $q_{(i)}$ are the Cartesian coordinates of
unit-mass point  particles. The single auxiliary $a$ matches the single
scaling degeneracy of the $q_{(i)}$. It  doubles the degeneracy. The
habitual pairing of $q_{(i)}$ and $a$ has several consequences, 
including full gauge invariance of all the theory's equations, as we
shall show elsewhere.  More importantly, it imposes constraints. Indeed,
the canonical momenta, 
$p^{(i)}=\partial\mathcal L/\partial\dot{q}_i$ and 
$p^{a}=\partial\mathcal L/\partial\dot a$, are
\beq
     p^{(i)} = \frac{\partial\mathcal L}
            {\partial\dot{\tilde q}_{(i)}}
        \frac{\partial\dot{\tilde q}_{(i)}}
           {\partial\dot{q}_{(i)}} 
        = \frac{\partial\mathcal L}
            {\partial\dot{\tilde q}_{(i)}}a,
    \hskip 1cm p^{a}  = \frac{\partial\mathcal L}
            {\partial\dot{\tilde q}_{(i)}}
        \frac{\partial\dot{\tilde q}_{(i)}}
            {\partial\dot a}
        = \frac{\partial\mathcal L}
            {\partial \dot{\tilde q}_{(i)}}q_{(i)}.
\end{equation}
We therefore have the {\it{primary constraint}} \cite{Dirac}
\beq 
    q_{(i)}p^{(i)}\equiv ap^{a}. \label{w1}
\end{equation}
The variation  $\int\textrm d\lambda(\delta\mathcal L/\delta a)\delta a$ is
\beq
    \int\textrm d\lambda
            \left(\frac{\textrm{d}p^{a}}
                    {\textrm{d}\lambda}
        - \frac{\partial\mathcal L}{\partial a}
            \right)\delta a 
        + p^{a}(\delta a)_{\textrm{final}} 
        - p^{a}(\delta a)_{\textrm{initial}}.
\label{w2}
\end{equation}
By the free-end-point rules, this must vanish for any $\delta a$.  
Starting with a general variation 
that vanishes at the end-points, we deduce the standard Euler--Lagrange equation
for $a$: 
$\textrm{d}p^{a}/\textrm{d}\lambda=\partial \mathcal L/\partial a$. Then, choosing a variation that does not vanish at the initial end-point but does vanish at
the final,  and enforcing the Euler--Lagrange equations, we deduce that 
$p^{a}_{\textrm{initial}} = 0$.  If we then choose 
a variation that does not vanish at the final end-point, we find that 
$p^{a}_{\textrm{final}} = 0$.  As we can freely 
choose the final point along the curve, we deduce that 
$p^{a}=0$.  The  Euler-Lagrange equation for $a$ reduces to 
$\partial \mathcal L/\partial a=0$.  This is a key {\it{consistency condition}}. Hence, one
can effectively  minimize the action with respect to $a$ and $\dot a$
independently, although the conditions really  arise from a
free-end-point variation of $a$.  Finally, we get a {\it secondary constraint} 
$q_{(i)}p^{(i)}=0$ from the primary constraint.  These simple considerations capture all
the novel  features of our theory.

We may now proceed to apply this variational procedure to the conformalized
BSW action~\eqref{action}.   However, we need one computation. We know
\beq
\wh{\phi} = {\phi [\int \sqrt{g} d^3x]^{1 \over 6} \over
              [\int \phi^6 \sqrt{g} d^3x]^{1 \over 6}}.
\end{equation}
Therefore
\beq
\dot{\wh{\phi}} = {\wh{\phi}\dot{\phi} 
\over      \phi}
-{\wh{\phi}\int \wh{\phi}^6 {\dot{\phi} \over \phi} \sqrt{g} d^3 x \over
\int\sqrt{g}d^3 x}
+ {\wh{\phi} \over 12} {\int ( 1 - \wh{\phi}^6)
g^{ab}\dot{g}_{ab}\sqrt{g} d^3 x \over
\int \sqrt{g} d^3 x}. \label{dotphi}
\end{equation}
  Now, variation with respect to
$\dot g_{ij}$ defines the canonical momenta
$\pi^{ij}$: 
\beq  
    \pi^{ij} \equiv \frac{\delta I}{\delta\dot{g}_{ij}} 
    = \Pi^{ij} 
    + \frac{<\tilde\Pi>}{3}\sqrt{g}\,g^{ij}(1 - \wh{\phi}^{6}),
\label{blah} 
\end{equation} 
where
\beq
    \Pi^{ij} = \frac{\wh{\phi}^2\sqrt{g}}{2N}
            G^{ijkl}\frac{d \wh{\phi}^{4}g_{kl}}{d\lambda},
    \hskip 1cm N \equiv \frac{1}{2}\sqrt{\frac{\wh{T}}
             {\wh{\phi}^{-4}(R - 8\frac{\grad^{2}\phi}{\phi})}},
\label{tarver}    
\end{equation}
and $\Pi = \sqrt{g}\,\tilde\Pi=g_{ij}\Pi^{ij}$.  Variation wrt
 $\dot\phi$ gives
\beq
    p_{\phi} \equiv \frac{\delta I}{\delta \dot{\phi}} 
        = \frac{4}{\phi}(\Pi - \sqrt{g}\,\wh{\phi}^6<\tilde\Pi>).
\label{pphi}\end{equation}
The `bulk' terms in Eqn.\eqref{blah} and Eqn.\eqref{pphi} come from
the second and third terms in Eqn.\eqref{dotphi}. These definitions
imply a primary constraint:
\beq
   p_{\phi} = \frac{4}{\phi}(\pi - \sqrt{g}<p>), 
\label{aumom}
\end{equation}
where $\pi = \sqrt{g} p = g_{ij}\pi^{ij}$.
Also, the rules of free-end-point variation imply that $p_\phi=0$.  
Hence we find the secondary constraint 
\beq
    p = C, 
\label{CMC1}
\end{equation} 
where $C$ is some spatial (ie $\lambda$-dependent) constant.
This constraint is equivalent to the CMC condition imposed by York, obtained here as
a genuine  constraint and not a gauge-fixing condition (a similar result
appeared earlier in~\cite{bom}).  As this condition designates a
preferred time-function, it introduces a notion of global simultaneity. 
This also means that our theory is restricted to those solutions of the
Einstein equations which possess a CMC slice.

Taking the trace of Eqn.\eqref{blah} 
\beq  
    \pi
    = \Pi 
    +<\tilde\Pi>\sqrt{g}(1 - \wh{\phi}^{6}),
\label{blah1} 
\end{equation} 
and integrating gives
\beq
 <p> = <\tilde\Pi> = C.
\label{blah2}
\end{equation}
When this is substituted back into Eqn.\eqref{blah1}, we get
\beq
 g_{ij}\Pi^{ij} = g_{ij}\frac{\wh{\phi}^2\sqrt{g}}{2N}
            G^{ijkl}\frac{d \wh{\phi}^{4}g_{kl}}{d\lambda} = 
C\sqrt{g}\wh{\phi}^{6},
\end{equation}
or
\beq
 -\frac{\wh{\phi}^{-4}g^{kl}}{N}
            \frac{d \wh{\phi}^{4}g_{kl}}{d\lambda} = C.
\label{CMC}
\end{equation}
This shows how $C$ must be a VPCT \eqref{VPCT} invariant. Such invariance
is a key part of the York construction.

The reparametrization-invariant momenta satisfy an identity analogous to 
the ADM Hamiltonian  constraint:
\beq 
    \pi^{ij}\pi_{ij} -\frac{\pi^2}{2} 
    - \frac{gC^2}{6}(1 - \wh{\phi}^{6})^{2} 
    + \frac{C}{3}\sqrt{g}\pi(1 - \wh{\phi}^{6}) 
     = g\wh{\phi}^{8}
        \biggl(R - 8\frac{\grad^{2}\phi}{\phi}\biggr).
\label{ham1}
\end{equation}
If we define 
$\sigma^{ij} \equiv \pi^{ij} - \frac{1}{3}g^{ij}\pi$ and enforce the CMC
constraint~\eqref{CMC1}, then~\eqref{ham1} becomes the identity
\beq
    \sigma^{ij}\sigma_{ij} - \frac{\pi^{2}\wh{\phi}^{12}}{6} 
        - g\wh{\phi}^{8}
        \biggl(R - 8\frac{\grad^{2}\wh\phi}{\wh\phi}\biggr)     = 0.
\end{equation}
It holds on any path in the configuration space.  In the Hamiltonian picture, in which $\pi^{ij}$ are
independent variables, it becomes a constraint.  It is in fact the {\it{Lichnerowicz--York equation}}~\cite{yor}.  Note that the conformal weighting of $\sigma^{ij}$ is exactly as imposed by Lichnerowicz and York. 

Varying wrt $\xi_{i}$ and using~\eqref{CMC1}, we obtain
\beq
    \grad_{j}\pi^{ij} = 0.\label{mom1} 
\end{equation}
This is the ADM momentum  constraint. In terms of our primary variables, it can be
written as
\beq
\grad_j\biggl[\sqrt{\frac
             {\wh{\phi}^{-4}(R - 8\frac{\grad^{2}\phi}{\phi})}{\wh{T}}}
     \wh{\phi}^2\sqrt{g}
            (g^{ik}g^{jl} - \frac{1}{3}g^{ij}g^{kl})\frac{d
\wh{\phi}^{4}g_{kl}}{d\lambda}\biggr] = 0.
\label{mom}
\end{equation}

Varying wrt $\phi$  
leads to the 
consistency condition
\beq 
   N \wh{\phi}^{-4} \left(R - 8\frac{\grad^{2}\phi}{\phi}\right) 
        - \wh{\phi}^{-6}\grad.\left(\wh{\phi}^2\grad{N} \right) 
        + \frac{N p^{2}}{4} = D,
\label{LFE}  
\end{equation} 
where we have used ~\eqref{CMC1} and
\beq
    D = \biggl<\wh{\phi}^{2}N
        \biggl(R - 8\frac{\grad^{2}\phi}{\phi}\biggr) 
        + \frac{\wh{\phi}^6 N p^{2}}{4}\biggr>;
\end{equation}
$D$ is the global average of the 
left-hand side of~\eqref{LFE}.

We now make a dimensional analysis. We take $g_{ij}$ and $\phi$ to be 
dimensionless and give dimensions to the coordinates. We set the speed of light equal to one, so all coordinates have dimensions of length (\textit{l}).  Counting derivatives 
in their definitions, we find the dimensions of other objects: 
$[R] = [\wh R] = [\wh T] = 1/l^2$.  The lapse, $N$, is dimensionless.  It follows that 
         $[\pi^{ij}] = [\pi] = [p] = [C] = 1/l$.  In the York scheme, one fixes an implicit length scale for the CWB manifold by specifying a numerical
value for the dimensionful parameter $C$.  That York's scheme determines a physical 3-metric from an initially  unphysical one reflects this fixing, as the manifold
acquires a volume corresponding to the implicit length scale.

Our approach is complementary to this. We specify as initial data
($g_{ij},  \dot{g}_{ij}$). Our procedure subjects them to considerable `gauge
dressing', but the initial volume $V$ corresponding to the initial
$g_{ij}$ is the one thing that must not be changed. We put the scale
in explicitly through the canonical coordinate $g_{ij}$, not though a
combination of the canonical momenta. In both cases, giving
the numerical value of a single quantity with a length dimension fixes
the values of all other quantities with a length dimension. The
time label is fixed only up to global reparametrization (as~\eqref{LFE}
is homogeneous in $N$), and, as always, the spatial coordinates are
freely specifiable.

Our calculations have given us five equations---the CMC 
condition~\eqref{CMC}, the momentum
constraint~\eqref{mom},  and the consistency condition~\eqref{LFE}---that
we try to solve for the five gauge variables 
$(\wh\phi, \partial\wh\phi/\partial\lambda, \xi_i)$ given the initial 
$g_{ij}, \dot{g}_{ij}$.  In accordance with our dimensional analysis the
constant $C$, with its dimension $l^{-1}$, emerges as an 
eigenvalue for this problem.  

There is an easy way of seeing this. Let us consider the three key
equations, Eqs.\eqref{CMC}, (27), (28), but with the `hats' removed from
the
$\phi$'s. Pick a value of $C$, and presume that the equations can be
solved for the ($\phi, \dot{\phi}, \xi$). Now consider the following
transformation
\beq
\phi \rightarrow \alpha \phi,\hskip 0.5cm \dot{\phi} \rightarrow \alpha
\dot{\phi},\hskip 0.5cm
\xi \rightarrow \xi,\hskip 0.5cm C \rightarrow \alpha^{-2} C,
\end{equation}
where $\alpha$ is any constant. This also generates a solution to the
three equations. Now, by choosing $\alpha = <\phi^6>^{-1/6}$, we can
normalize $\phi$ and thus get the desired solution. In the process we fix
the appropriate value of $C$.

There are many choices of $(g_{ij},
\dot g_{ij})$ for which we know  solutions exist; e.g.~take a CMC slice
through a space-time that solves the Einstein equations,  take the
physical $g_{ij}$ and $\dot g_{ij}$ given by the CMC foliation, and
multiply by any VPCT.   On the other hand, with a given arbitrary
$(g_{ij}, \dot g_{ij})$, we have no existence theorem for  solutions. 
This problem is very similar to the thin-sandwich problem in
GR~\cite{bsw, ts}.  As in standard canonical general relativity, the constraints are much easier to solve in the Hamiltonian picture than in the action. The key equation is a slight modification of the standard Lichnerowicz-York equation. It has been shown, \cite{LY}, that this equation has extremely nice existence and uniqueness properties.

If and when we can solve the equations, we can switch to the `corrected
coordinates'  $\overline g_{ij}$.  In this frame, $\wh\phi = 1$, and the consistency
condition~\eqref{LFE} reduces  to
\beq
    N R - \nabla^2 N + \frac{N p^2}{4} = D.
\end{equation}
This is the CMC lapse-fixing equation of GR; it ensures propagation of
the  CMC constraint~\eqref{CMC1}.

York retained  the spacetime ontology and used conformally invariant decoupling of the ADM Hamiltonian and momentum constraints  as a guiding principle. This led him to regard the Lichnerowicz--York equation as a gauge fixing  (see especially his
comments near the end of the first paper in~\cite{yor}).  In contrast, our ontology resides in CS+V, and our initial-value problem contains a  genuine gauge invariance. This leads to gauge corrections to both $g_{ij}$ and $\dot{g}_{ij}$.  Our new principles enable us to derive Hamiltonian GR, the prescription for solving its initial-value
problem, and the condition for maintaining the  CMC condition in a single
package. The details of the transition to  the Hamiltonian formulation
and the implications of our results will be considered elsewhere.

Three final remarks. First, our work shows that to derive the dynamical
core of GR one can use either Einstein's `relativity of simultaneity' or
our `relativity of local size' but not both at once.  The relativity of
size approach is clearly very powerful since it yields not only GR but
also the method of solving its difficult initial-value problem. Second,
our analysis could naturally be called  `the conformal thin-sandwich'
approach to GR. Unfortunately, this phrase has already been used
\cite{cts}. The relationship between the two approaches deserves further
investigation. Third, we obtain GR in the CMC gauge. This covers a very
large class of solutions of the Einstein equations, but not all, see
\cite{i}. This is a problem that will be associated with any
gauge-fixing. Whether this is to be regarded as a `good' or a `bad'
feature of our approach, only time will tell.

\begin{acknowledgments}
We dedicate this paper to Jimmy York on the occasion of his 65th
birthday.  The authors thank him for discussions. 
EA also thanks Malcolm MacCallum for discussions and PPARC for financial
support.  EA, BF, BK, and N\'OM thank the Barbour family for hospitality.
EA acknowledges funding from Peterhouse and from the Killam Foundation, and thanks Queen Mary University of London for Visitor Status. BF was supported in part by the NSF under grant PHYS-0300710 at the 
University of Maryland and by the CNRS at the Institut d'Astrophysique de Paris. BK thanks Bill Unruh and Johan Br\"annlund for discussions.

\end{acknowledgments}


\begin{thebibliography}{29}

\bibitem{adm}   
R.~Arnowitt, S.~Deser and C.~Misner 
in {\it Gravitation: an Introduction to Current Research}, 
ed. L.~Witten, (Wiley, New York, 1962), arXiv:gr-qc/0405109;
C.~Misner, K.~Thorne, and J.~A.~Wheeler, {\it{Gravitation}} (Freeman,
San Francisco, 1973).

\bibitem{yor}
J.~W.~York,
Phys.\ Rev.\ Lett.\  {\bf 26}, 1656 (1971);
J.~W.~York,
Phys.\ Rev.\ Lett.\  {\bf 28}, 1082 (1972);
J.~W.~York, 
J.\ Math.\ Phys.\ {\bf 14}, 456 (1973). 


\bibitem{Lich}  
A. Lichnerowicz, J.\ Math.\ Pures\ Appl.\ {\bf 23}, 37 (1944).


\bibitem{CS}    
J.~W.~York, 
Ann.\ Inst.\ Henri Poincar\'{e} {\bf 21}, 319 (1974); 
A.~E.~Fischer and V.~Moncrief,
Gen.\ Rel.\ Grav.\  {\bf 28}, 221 (1996).

\bibitem{bsw}   
R.~F.~Baierlein, D.~Sharp and J.~A.~Wheeler, 
Phys.\ Rev.\ Lett.\ {\bf 126}, 1864 (1962).

\bibitem{rwr}
J.~Barbour, B.~Z.~Foster and N.~\'O~Murchadha,
Class.\ Quant.\ Grav.\  {\bf 19}, 3217 (2002)
E.~Anderson,
Phys.\ Rev.\ D {\bf 68}, 104001 (2003)
E.~Anderson, PhD Thesis: {\it Geometrodynamics: Spacetime or Space?}, 
Univ. London (2004). 

\bibitem{bom}
J.~Barbour and N.~\'O~Murchadha,
arXiv:gr-qc/9911071;
E.~Anderson, J.~Barbour, B.~Foster and N.~\'O~Murchadha,
Class.\ Quant.\ Grav.\  {\bf 20}, 1571 (2003)
B.~Kelleher,
Class.\ Quant.\ Grav.\ {\bf 21}, 483 (2004); {\bf 21}, 2623 (2004).

\bibitem{jb}    
J.~Barbour,
Class.\ Quant.\ Grav.\  {\bf 20}, 1543 (2003)

\bibitem{Dirac} 
P.~A.~M. Dirac, 
{\it Lectures on Quantum Mechanics}, (Yeshiva University, New York,
1964).

\bibitem{LY}
N. \'O Murchadha, Acta Physica Polonica B, {\bf 36}, 109 – 120 (2005); gr-qc/0502055.

\bibitem{ts}    
E.~P.~Belasco and H.~C.~Ohanian, 
J.\ Math.\ Phys.\ {\bf 10}, 1053 (1969); 
R.~Bartnik and G.~Fodor,
Phys.\ Rev.\ D {\bf 48}, 3596 (1993)
D.~Giulini,
J.\ Math.\ Phys.\  {\bf 40}, 2470 (1999)

\bibitem{cts}
J.~W.~York, Phys.\ Rev.\ Lett. {\bf 82}, 1350 (1999);
H.~Pfeiffer and J.~W.~York, Phys.\ Rev. {\bf D 67}, 044022 (2003);
J.~W.~York, arXiv:gr-qc//0405005.

\bibitem{i}
R. Bartnik, Comm.\ Math.\ Phys. {\bf 117}, 615 (1988); D. Eardley and D.
Witt, unpublished (1992); P. Chrusciel, J. Isenberg, and D. Pollack,
Phys.\ Rev.\ Lett., {\bf 93}, 081101 (2004).

\end{thebibliography}
\end{document}